\documentclass[runningheads]{llncs}
\usepackage[T1]{fontenc}
\usepackage{graphicx}
\usepackage{hyperref}
\usepackage{tabularx}
\usepackage{amsmath}
\usepackage{float}
\usepackage{subcaption}

\begin{document}
\title{Changes in Help-Seeking Strategies Predict unaided Performance during AI-based Mathematical Learning}
\titlerunning{AI Interactions in Mathematical Education}

\author{Rania Abdelghani\inst{1}\orcidID{0000-0002-6361-6609} \and
Peter Kaiser\inst{1,2}\orcidID{0000-0003-1031-1139} \and
Kou Murayama\inst{1}\orcidID{0000-0003-2902-9600 }}

\authorrunning{R. Abdelghani et al.}

\institute{Hector Institute of Education Sciences and Psychology, University of Tübingen, Germany\\
\email{\{rania.abdelghani,pe.kaiser,k.murayama\}@uni-tuebingen.de}
\and
Department of Mathematics, Faculty of Sciences, University of Tübingen, Germany\\
}
\maketitle              
\begin{abstract}
Generative AI systems are increasingly used by students as learning companions, yet little is known about \textit{how} they regulate these tools in open-ended learning settings, where the goal is not to complete a predefined task but to improve understanding for later independent performance. This study examined Grade-9 students' use of a general-purpose LLM during curriculum-aligned mathematical-modeling practice through the lens of \emph{epistemic proactivity (EP)}---the extent to which learners retain responsibility for regulating and advancing their own knowledge-building while using AI. We operationalized EP through two complementary behavioral dimensions: self-regulated learning (SRL) functions and help-seeking (HS) content. We further asked whether static summaries of these behaviors or their within-session development were more informative about subsequent unaided performance. A total of 112 students interacted with a web-based LLM tutor, producing 880 substantive student turns; 95 completed both the unaided pre- and post-tests. Interactions were strongly request-dominated, with comparatively little explicit planning, monitoring, or evaluation, and students' stated HS intentions were only weakly reflected in their enacted behavior. Static summaries of AI use did not account for additional post-test variance beyond pre-test performance. In contrast, among the 89 students with HS computable trajectories, within-session HS development was informative: students whose requests became relatively more understanding-oriented---toward conceptual and procedural assistance rather than answer-seeking or verification---showed higher subsequent unaided performance. These findings suggest that AI-supported learning cannot be understood from aggregate interaction characteristics alone; the change in the depth with which students formulate their requests as an interaction develops may provide additional insight into their subsequent performance without AI.
\end{abstract}
\label{sec:intro}
Generative AI (GenAI) systems are increasingly becoming part of students' everyday learning practices~\cite{hashem2025understanding}. Their educational promise lies in providing conversational and on-demand support: learners can request explanations, feedback, or worked solutions whenever they feel stuck~\cite{kasneci2023,holmes2023guidance}. In mathematics, such flexibility can be particularly useful because learners must coordinate conceptual knowledge, procedures, representations, and real-world information while solving problems~\cite{bulut2026ai}. Yet easier access to assistance does not necessarily translate into learning that remains once AI support is removed~\cite{Yan2025DistinguishingAI}. Without proper control, LLMs can perform parts of the reasoning that learners would otherwise need to carry out themselves, allowing immediate task success while bypassing the processes and struggles through which understanding is constructed~\cite{stadler2024cognitive,shaw2026thinking}.

We use the term \emph{epistemic proactivity (EP)} to characterize the extent to which learners retain this responsibility to carry out the important process for learning. This concerns not simply whether learners are actively seeking assistance, but \emph{what role} they retain during such a process: are they the ones framing problems, are they the ones originating learning trajectories or simply reacting to ones imposed by the AI, and so on~\cite{yan2026agentivism}. EP consequently goes beyond prompt sophistication to understand students' learning efficiency with AI, or general behavioral engagement as broad constructs: it focuses specifically on how responsibility for knowledge-building is distributed between learner and AI.

Two long-standing learning processes provide complementary behavioral lenses to operationalize EP. First, self-regulated learning (SRL) concerns how learners plan goals and strategies to advance learning, monitor their understanding, seek or enact strategies to reach their goals, and evaluate outcomes in order to adapt subsequent behavior~\cite{Zimmerman2008InvestigatingProspects}. In AI-supported learning, explicit planning, monitoring, and evaluation can therefore provide evidence that learners themselves contribute to regulating the interaction, rather than acting in a purely reactive way to directions introduced by the AI~\cite{gerlich2025ai}. Second, help-seeking (HS) characterizes \emph{what assistance} learners ask an external source to provide in order to reach their epistemic goal~\cite{nelson1986help}. Conceptual and procedural requests can support continued reasoning and understanding, whereas answer- and verification-seeking provide more immediate epistemic closure by asking the external source to provide or confirm an outcome~\cite{graesser1994question}. Importantly, these forms of assistance are not inherently productive or unproductive for learning; their epistemic value depends on how they are organized and adapted to the learner's current goals, task requirements, and available cognitive resources~\cite{Seufert2018TheLoad}.

This qualification is particularly important because productive learning does not require learners to perform every cognitive and regulatory operation unaided. Assistance can reduce demands and allow resources to be allocated to other relevant aspects of the task. The central problem is therefore one of calibration: support should facilitate progress without progressively replacing the learner activity through which learning occurs~\cite{yan2026agentivism}. With LLMs, this balance may be particularly difficult to maintain because the system can not only execute requested operations, but also take over planning, monitoring, and evaluation, or suggest how the interaction should proceed (i.e. what next requests should be). As learners experience such delegation without an apparent cost, progressively offloading these EP-related processes may become increasingly attractive~\cite{gerlich2025ai}. EP should therefore be understood as a developing property of the interaction rather than as a fixed characteristic of either the learner or an isolated prompt.

Yet much research on educational GenAI still characterizes use through surveys, usage frequency, or whole-session behavioral distributions~\cite{liu2026behavioral}. Emerging process-sensitive work has begun to examine sequential and temporal organization in student--AI interaction~\cite{alamray2026exploring}, but evidence remains limited regarding how within-session behavioral development relates to subsequent performance once AI assistance is removed. This work is also highly concentrated on university students.

We suggest that reduced EP may be particularly consequential for younger learners, who are still developing the cognitive, metacognitive, and regulatory capacities needed to maintain epistemic responsibility while learning with AI~\cite{ruggeri2016sources}. Existing studies indeed suggest that secondary-school students often rely on relatively simple AI requests and do not consistently evaluate or interpret AI-generated information~\cite{bulut2026ai,abdelghani2025investigating}. Examining their actual conversational behavior is therefore especially important for identifying where additional support may be needed, particularly as adolescents' use of general-purpose GenAI tools for learning continues to increase~\cite{maheux2026generative}.

\section{Current Study and Expected Contributions}
The present study examines EP in the context of Grade-9 Mathematical modeling (MM) practice with a general-purpose LLM. MM provides a particularly informative context because learners must not only apply mathematical knowledge, but also understand and structure real-world situations, select and execute mathematical procedures, interpret results, and evaluate their validity~\cite{blum2009mathematical,ferri2010mathematical}. Moreover, students in our study were not asked to use the LLM to solve a predefined problem. Instead, they were instructed to use the interaction to advance their understanding of MM in preparation for a later assessment. Decisions about what they still needed to understand, how to organize the interaction, and what assistance to seek were therefore largely left to the students. This open-ended setting makes it possible to examine learners' spontaneous development of EP while using AI.

Building on the framework developed above, we examine EP through students' SRL functions and the epistemic orientation of their HS requests. We first characterize the explicit regulatory functions expressed in students' dialogue and the forms of assistance they seek, including whether their stated pre-interaction goals are reflected in their subsequent HS behavior. We then examine how these behaviors change within the session, rather than considering only whole-session distributions. Finally, we test whether these within-session changes are associated with subsequent unaided post-test performance beyond pre-test performance and static characteristics of AI use.

Accordingly, we address three research questions:
\begin{enumerate}
    \item \textbf{RQ1:} How do students regulate their interaction with the LLM and what forms of help do they seek during open-ended mathematical practice?
    \item \textbf{RQ2:} How do students' SRL and HS behaviors change over the course of the interaction?
    \item \textbf{RQ3:} Do within-session changes in these behaviors account for subsequent unaided post-test performance beyond pre-test performance and static patterns of AI use?
\end{enumerate}

The study contributes a process-oriented account of AI-supported learning that complements approaches based on isolated prompts or whole-session usage summaries. Rather than assuming that particular request types are intrinsically productive or unproductive, we examine whether the distribution of epistemic responsibility changes as students interact with the AI. In doing so, the study aims to provide first evidence about the value of considering the temporal organization of student--AI interaction when evaluating and supporting AI use in secondary-school learning.

\section{Methods}
\label{sec:methods}
\subsection{Participants}
We recruited 112 Grade-9 students from three public German \textit{Gymnasium} schools, aged 14--15 ($M=14.67$). Only 95 had completed both the pre- and post-AI interaction assessment.

The study was approved by the institute's ethics committee and the Ministry of Culture, Youth and Sports. Written informed consent was obtained from students and their legal guardians.

\subsection{Study Material and Procedure}
Students completed a three-phase tablet-based protocol during a regular mathematics lesson. After a short presentation by the research team, they completed a mathematics test with three validated linear and quadratic items selected for curricular fit, used to assess prior skills. They then interacted with Mistral Large 3 on a real-world MM skill grounded in three worked examples.

\begin{figure}[ht]
\centering
\includegraphics[width=1\textwidth, clip]{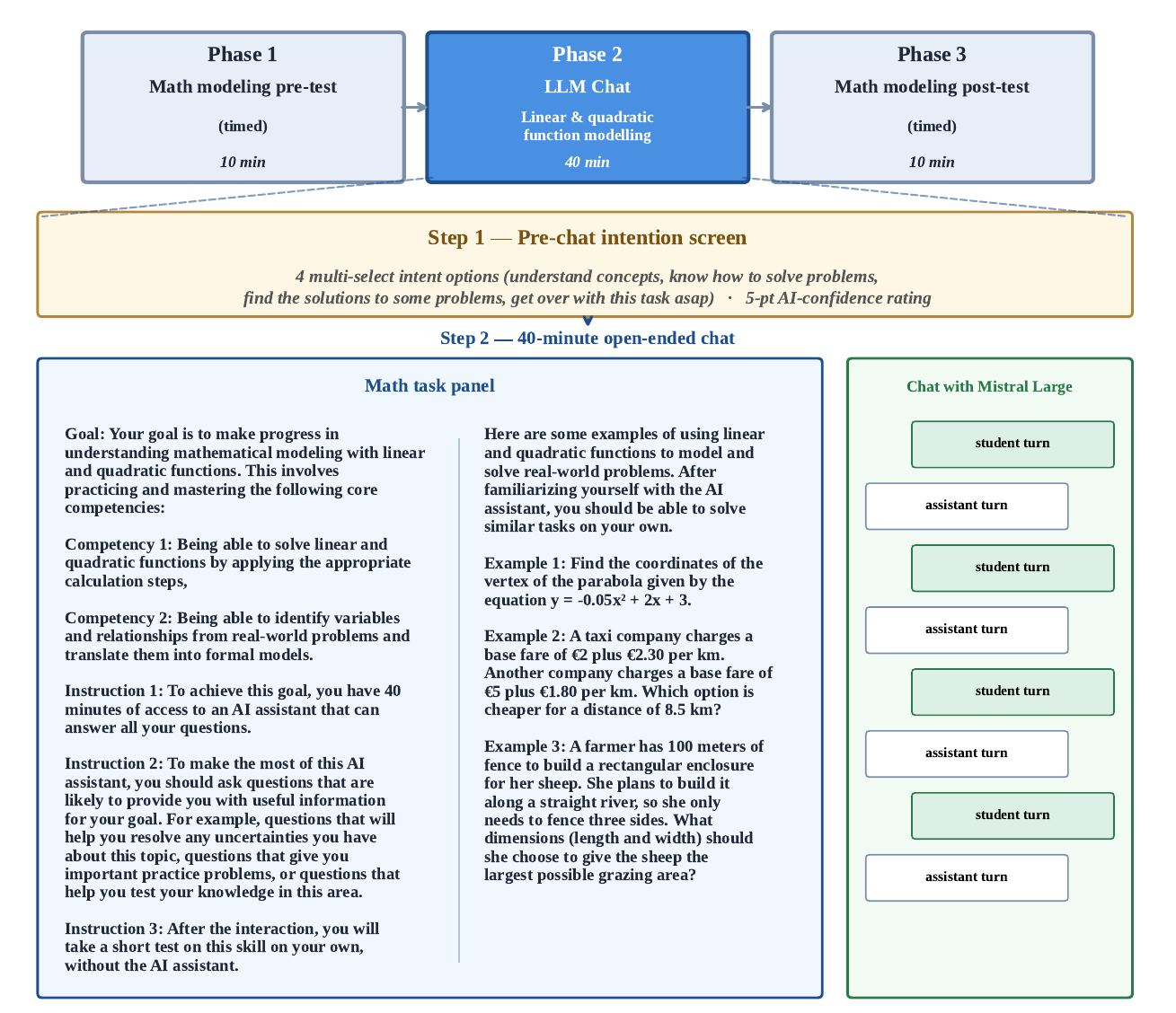}
\caption{Timeline of the study and content of the pedagogical task} 
\label{fig:stud_time_content}
\end{figure}

The AI-supported activity was framed as preparation for a later assessment rather than completion of a predefined problem. Students were asked to use the LLM to advance their understanding of mathematical modelling by addressing uncertainties, requesting practice, clarifying concepts, or working through difficulties. They were given the target competencies and examples of ways in which the AI could be used, but no prescribed sequence of activities or specific problem that they had to solve. Decisions about what to address, how to organize the interaction, and what assistance to request were therefore largely left to the students. See~\autoref{fig:stud_time_content}.

The complete instructions, task materials, and interaction interface are provided in the supplementary material, available in~\footnote{\label{smfootnote}\url{https://osf.io/p4y8c/overview?view_only=1b23058abb4a4b48931d939a6a0e3147}}.

After reading the task and before starting the chat, students selected the learning goals they wanted to pursue with the AI assistant, from options inspired by the questions taxonomy explained above~\cite{graesser1994question,tawfik2020role}: wanting to understand concepts and explanations, wanting procedures to know how to solve problems and practice, want to solve specific problems, or finish the task as soon as possible (amotivation). These selections were subsequently used to compare students' stated pre-interaction intentions with their enacted HS behavior, as an addition potential SRL indicator.

Students then interacted freely with the LLM. To ensure that each participant engaged sufficiently with the activity for conversational analysis, they were required to produce at least six task-relevant student turns before completing the chat phase. Off-task or uninterpretable turns were not counted toward this minimum and were excluded from the substantive conversational corpus. After the AI-supported phase, students completed the cognitive-load questionnaire and the unaided post-test described below.

\subsection{Measures}
\label{sec:measures}
\subsubsection{Pre- and Post-Interaction Mathematical tests} Students' MM skills were assessed with pre- and post-test tasks drawn from the IQB VERA mathematics item pool. The selected tasks targeted MM competencies and were chosen using the official IQB metadata on competence area, content focus, and difficulty level. We used different items at pre- and post-test to avoid item-repetition, memory, or direct practice effects. Pre- and post-test items targeted the same MM concepts and were matched on difficulty level according to the official IQB classification. In the analyses, pre-test performance was used as an estimate of students' prior MM skill and included as a covariate when predicting AI-free post-test performance. Performance in both tests was annotated manually by the research team.

\subsubsection{Conversational Data: Codebook Development, Validation, and Application}
We developed a theory-driven turn-level coding framework to characterize students' EP during interaction with the LLM. Only substantive, task-relevant student turns were assigned behavioral codes. Epistemic proactivity was operationalized through two complementary dimensions corresponding to the framework developed~\autoref{sec:intro}: the regulatory function expressed in the turn and, for requests, the epistemic orientation of the assistance sought.

First, each substantive student turn was coded for its explicit SRL function, drawing on Zimmerman's cyclical model~\cite{Zimmerman2008InvestigatingProspects}. \textsc{Plan} captured explicit structuring of goals or strategies; \textsc{Monitor} captured expressions of current understanding, uncertainty, or knowledge gaps; \textsc{Request} captured turns in which the student sought assistance from the AI; and \textsc{Evaluate} captured explicit appraisal of the AI's response or of progress toward the learning goal. These codes describe regulatory functions expressed explicitly in the dialogue and should not be interpreted as a complete measure of students' internal self-regulation. 

Second, turns coded as \textsc{Request} were further classified according to the assistance sought~\cite{nelson1986help,graesser1994question}. \textsc{Conceptual} requests sought explanations of ideas or relations, \textsc{Procedural} requests sought support for how to perform or approach a task, \textsc{Verification/ validation} requests asked the AI to assess or confirm existing reasoning or an answer, and \textsc{Answer-seeking} requests asked the AI to provide the outcome directly. For the temporal analyses described below, conceptual and procedural requests were contrasted with answer- and verification/validation-seeking requests. This contrast does not assume that either set of behaviors is intrinsically productive or unproductive; rather, it distinguishes requests that primarily support the learner's continued construction of understanding from those that ask the AI to provide or confirm epistemic closure. Separately, we annotated MM activity using the modeling cycle in~\cite{blum2009mathematical}. These codes were used to describe the domain-specific content of students' interactions, rather than to define epistemic proactivity, because stages such as interpretation and validation can represent productive mathematical activity depending on their role in the ongoing task.

The codebook was developed iteratively. Two trained human coders independently applied a preliminary version to a stratified subset of 198 turns (approximately 20\% of the original corpus). Initial agreement was substantial for SRL function ($\kappa=.76$)) but lower for HS content ($kappa=.53$)) and MM activity ($\kappa=.60$)). Disagreements were used diagnostically to clarify category definitions, introduce ordered decision rules, and refine examples of clear and borderline cases, after which the codebook was frozen.

Gemini 3.1 Pro was then used to apply the resulting human-developed framework at scale, with each target turn interpreted in the context of the complete conversation. For development and evaluation of the automated annotation procedure, the 198 human-coded turns were divided into three non-overlapping sets: 42 clear agreement cases were used as in-context examples fed to the model, 30 additional cases were used to refine the annotation instructions, and 126 cases were held out from this automated-annotation development process for final evaluation. A new expert applied the frozen codebook to these 126 cases while blinded to the LLM annotations, producing expert-resolved reference labels. Agreement between these labels and the automated annotations reached $\kappa=.75$ for SRL function and $\kappa=.65$ for HS content. MM classification reached only $\kappa=.41$ and was therefore excluded from subsequent analyses. The validated procedure was then applied to the full corpus of 880 substantive student turns. The complete codebook, prompts, model settings, examples, and validation materials are provided in the supplementary material.

Because the temporal analyses relied on collapsed contrasts, we additionally evaluated reliability for the exact distinctions used in these analyses. For HS, conceptual and procedural requests were collapsed into an understanding-oriented category and verification- and answer-seeking requests into a closure-oriented category, yielding substantial agreement with the expert-resolved labels ($\kappa=.84$)). For SRL, \textsc{Request} was contrasted with the combined \textsc{Plan}, \textsc{Monitor}, and \textsc{Evaluate} category, yielding ($\kappa=.70$).

\section{Results}
\label{sec:results}
\subsection{Sample Description}
The interaction corpus comprised 988 student turns from 112 students. After excluding off-task and non-substantive input, 880 turns were retained for the conversational analyses ($M=7.86$, $SD=5.84$ substantive turns per student). 95 students completed both the unaided pre- and post-tests and were retained for analyses involving post-test performance. Descriptive statistics are reported in~\autoref{tab:measures}.

\begin{table}
\caption{Measures taken before, during, and after the AI interaction}
\label{tab:measures}
\centering
\renewcommand{\arraystretch}{1.4}
\begin{tabular}{|>{\raggedright\arraybackslash}p{0.34\textwidth}|c|c|c|c|}
\hline
Construct & $N$ & Mean & SD & Range \\
\hline
Pre-test Math Performance (\%)        & 95  & 68.04 & 31.15 & 0--100 \\
Total Time Spent in Chat (min)        & 112 & 28.60 & 6.17  & 0--40 \\
Mean AI-response-viewing Time (s)     & 112 & 148   & 81    & -- \\
Number of Substantive Student Turns   & 112 & 7.86  & 5.84  & -- \\
Post-test Math Performance (\%)       & 95  & 53.20 & 26.40 & 0--100 \\
\hline
\end{tabular}
\end{table}

\subsection{Overall Interaction Patterns}
\begin{figure}[ht]%
    \centering
    \subfloat[\centering \textbf{Distribution of Coded Students Turns} ]{{\includegraphics[width=.47\textwidth]{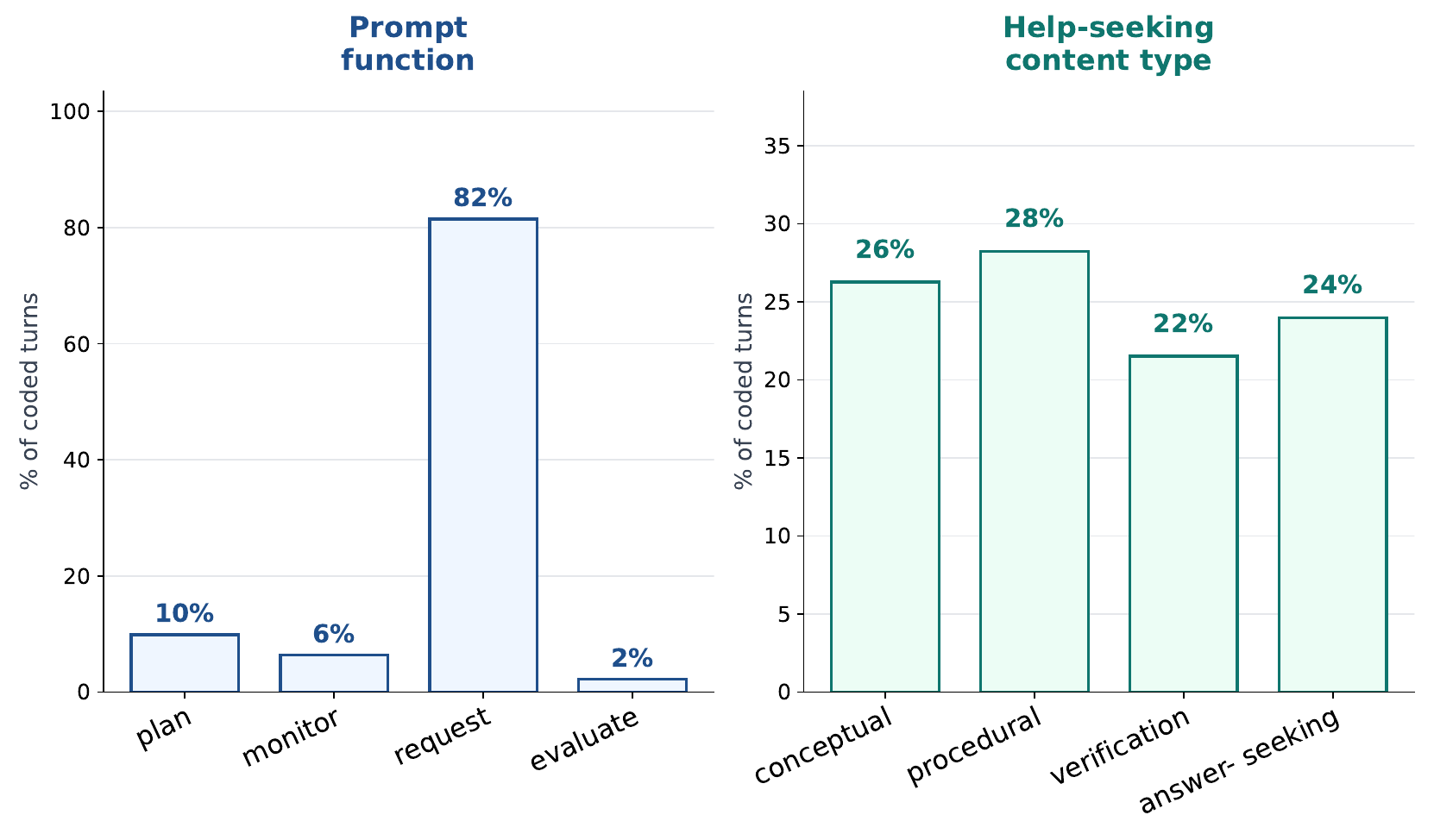} }}%
    \subfloat[\centering \textbf{Stated Intention vs. enacted Behavior}]{{ \includegraphics[width=.47\textwidth]{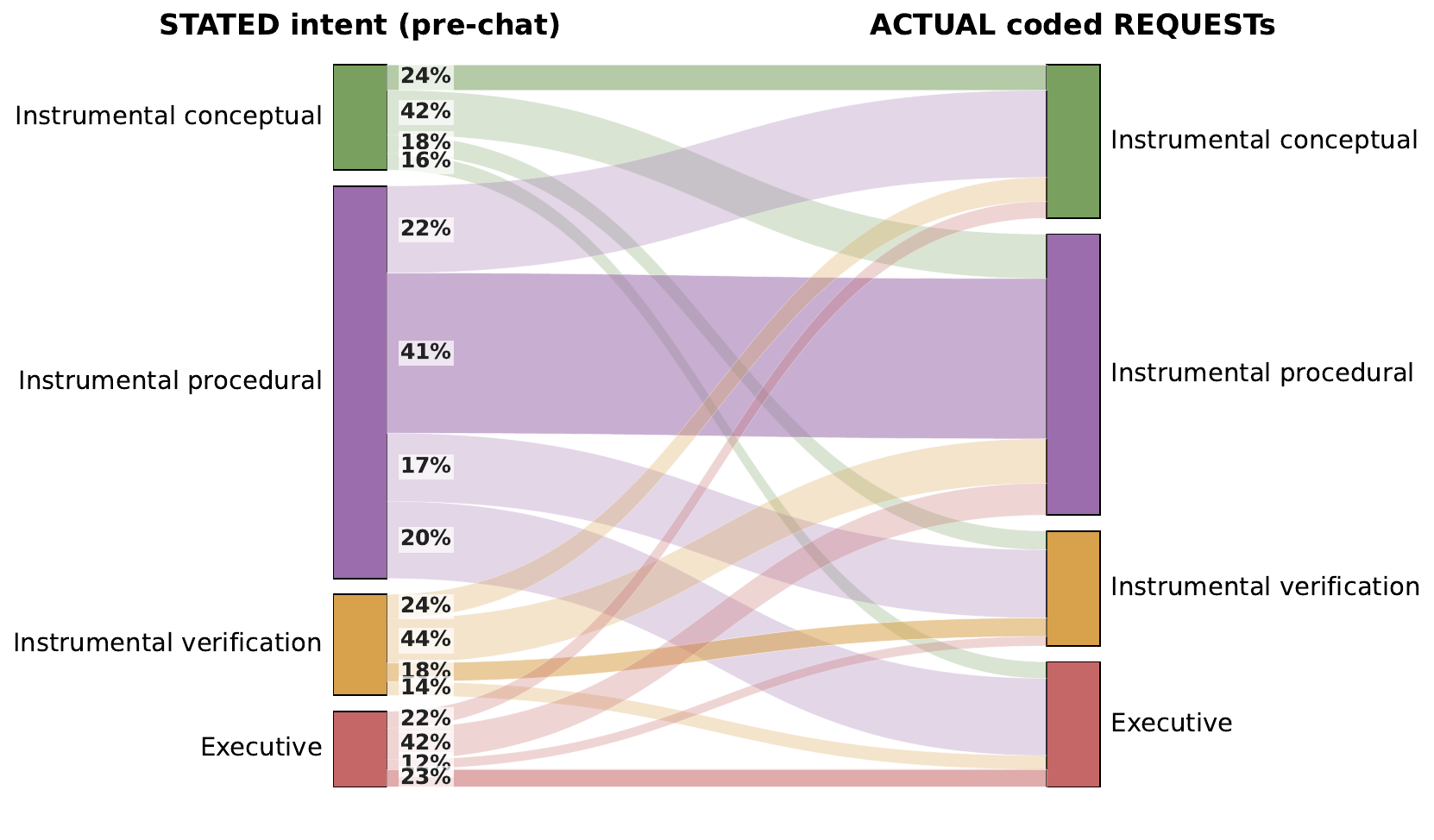} }}%
    \caption{\centering Description of students' LLM-supported mathematics interaction}%
    \label{fig:res-dist}%
\end{figure}

\paragraph{SRL Functions and Help-Seeking.}
We first examined the overall distribution of students' SRL functions and HS strategies. As shown in~\ref{fig:res-dist}, interactions were strongly dominated by requests for assistance. Of the 880 substantive student turns, 82\% were classified as \textit{REQUEST}, compared with 10\% \textit{PLAN}, 6\% \textit{MONITOR}, and 2\% \textit{EVALUATE}. Thus, although requesting assistance is itself part of self-regulation, explicit planning, monitoring, and evaluation were comparatively rarely expressed in students' dialogue.

The epistemic orientation of students' requests was more evenly distributed. Among the \textit{REQUEST} turns, 28\% were procedural, 26\% conceptual, 24\% answer-seeking, and 22\% verification-seeking. Understanding-oriented requests---conceptual and procedural assistance intended to support continued reasoning or problem solving---therefore accounted for 54\% of HS, whereas 46\% consisted of answer-seeking or verification. Students thus relied extensively on the tutor for assistance, but differed in whether they sought support for advancing their understanding or asked the AI to provide or confirm an outcome.

\paragraph{Stated Intentions and Enacted Help-Seeking.}
We next examined whether students' stated purposes for using the tutor before the interaction corresponded to the HS strategies they subsequently enacted. Overall correspondence between intended and enacted categories was weak: 32.9\% of the intention--behavior mass fell on matching categories, close to the 32.4\% expected from their marginal distributions. The largest discrepancy concerned procedural support, which was more prominent in students' stated intentions than in their enacted HS. Conversely, conceptual and answer-seeking requests occurred somewhat more frequently in the interaction than students had initially indicated. Thus, students' stated intentions provided only limited information about how they subsequently used the LLM. See~\autoref{fig:res-dist}.

\subsection{Within-Session Development}
\begin{figure}%
    \centering
    \subfloat[\centering \textbf{Overall Changes in SRL Functions and HS strategies across Normalized Session Position.} ]{{\includegraphics[width=.47\textwidth]{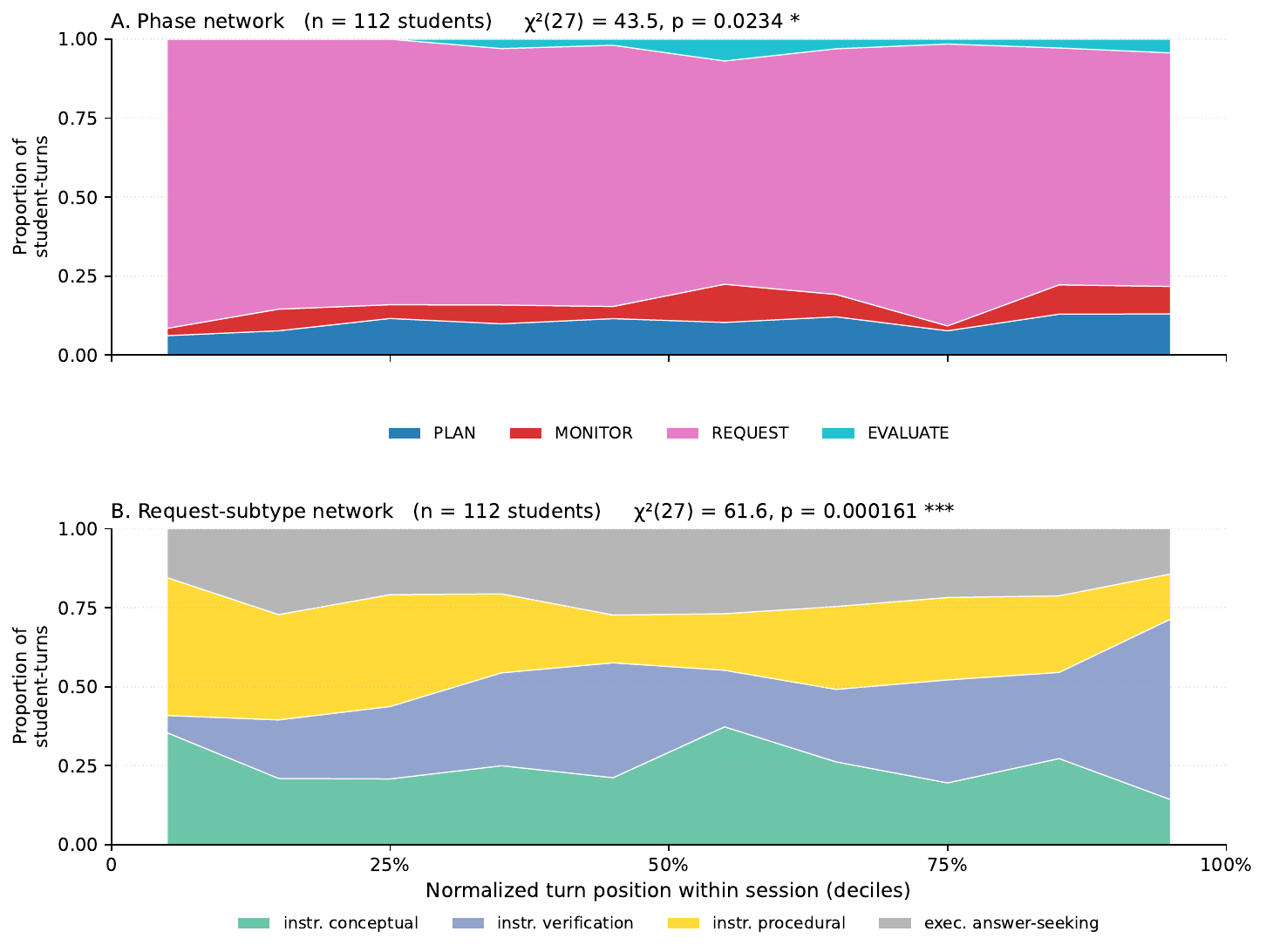} }}%
    \subfloat[\centering \textbf{Distribution of SRL and HS Temporal Change Indices}]{{ \includegraphics[width=.47\textwidth]{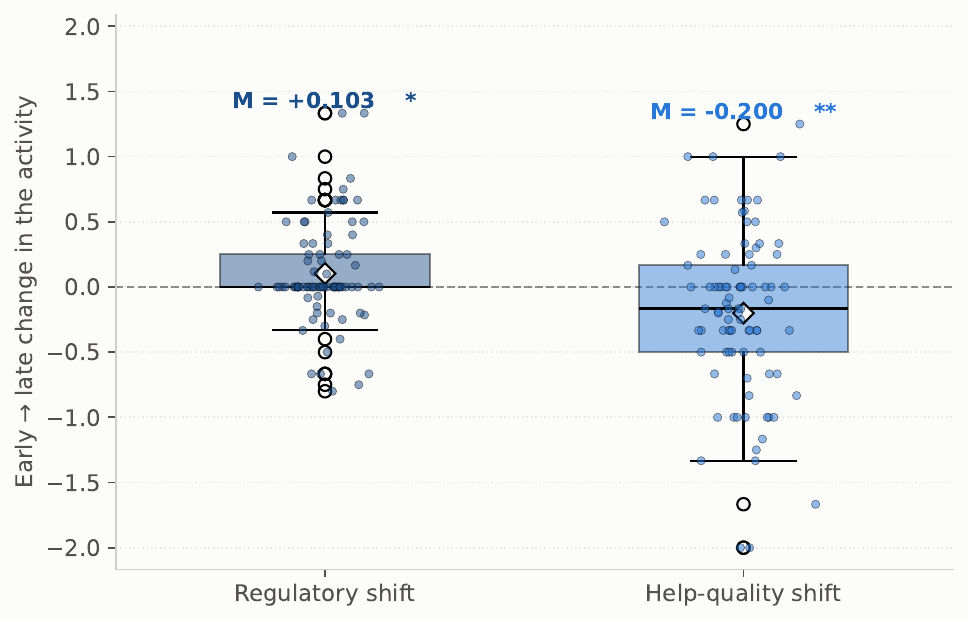} }}%
    \caption{\centering Within-session development of students' SRL functions and HS strategies.}%
    \label{fig:temp_descriptions}%
\end{figure}

We first examined whether students' SRL and HS changed as the interaction progressed. To account for repeated turns nested within students, we estimated turn-level mixed-effects models using normalized within-session position. At the cohort level, the two dimensions developed in different directions. HS shifted away from conceptual and procedural assistance and toward answer- and verification-seeking as the interaction progressed (population slope $=-.40$, $p<.001$). In contrast, SRL functions shifted modestly away from request-dominated interaction toward greater explicit planning, monitoring, or evaluation (population slope $=.26$, $p<.001$). Thus, students expressed somewhat more explicit regulation over time while simultaneously becoming less understanding-oriented in the assistance they requested. See~\autoref{fig:temp_descriptions}

To examine whether students' epistemic behavior changed over the interaction, each substantive turn was assigned a normalized within-session position ranging from 0 to 1. For each student, turns before the midpoint ($t<.5$) were classified as early and those at or after the midpoint ($t\geq.5$)) as late. For SRL, we computed the early-to-late change in the balance between explicit regulatory functions (\textit{PLAN}, \textit{MONITOR}, \textit{EVALUATE}) and \textit{REQUEST}, reflecting on the extent to which students drifted towards more explicit regulation over time. For HS, we analogously computed the change in the balance between understanding-oriented requests (conceptual, procedural) and more closure-oriented requests (verification, answer-seeking).

For each dimension $d$, the trajectory index was defined as:
\medskip

\noindent\makebox[\linewidth][c]{%
$\displaystyle
\Delta_d =
\left[
p_{\mathrm{late}}(O^{+}_d) - p_{\mathrm{late}}(O^{-}_d)
\right]
-
\left[
p_{\mathrm{early}}(O^{+}_d) - p_{\mathrm{early}}(O^{-}_d)
\right]
$
}

\medskip

where $O^{+}_d$ and $O^{-}_d$ denote the two theoretically contrasted orientations within each dimension. For SRL, $O^{+}$ comprised \textit{PLAN}, \textit{MONITOR}, and \textit{EVALUATE}, and $O^{-}$ comprised \textit{REQUEST}. Positive scores therefore indicate increasing explicit regulation for SRL or increasingly understanding-oriented assistance for HS; negative scores indicate the opposite. These scores capture coarse directional change across the session rather than complete individual trajectories.

At the student level, these change scores showed the same contrasting pattern: the SRL change was modestly positive ($M=.10$, $SD=.39$), whereas the HS change was negative ($M=-.20$, $SD=.64$), indicating that students tended to express somewhat more explicit regulation while their requests became relatively more closure-oriented over the interaction. Importantly, these two developments were largely independent: becoming more explicitly regulatory did not systematically imply becoming more understanding-oriented in HS.

\subsection{Links with Subsequent Unaided Performance}
We next tested whether subsequent unaided post-test performance was better accounted for by students' overall interaction patterns or by how these patterns changed within the session. These analyses used the 95 students with complete pre- and post-test data and computable temporal indicators. However, six more students with complete assessments could not receive an HS change score because one session half contained no \textsc{Request} turns. All models therefore used the same matched final sample, $N= 89$.

We first examined four order-independent, whole-session characteristics of AI use. Two described interaction composition: the proportion of SRL turns classified as \textit{PLAN}, \textit{MONITOR}, or \textit{EVALUATE}, and the proportion of requests classified as conceptual or procedural. Two additional overall metrics indicators captured interaction amount and pace: total chat time and mean AI-response review time. Pre-test performance alone accounted for $R^2=.161$ of post-test variance. Adding these static interaction features increased explained variance only to $R^2=.184$, $\Delta R^2=.023$, $F(4,83)=0.57$, $p=.684$. Thus, students' overall interaction composition, time spent interacting with the AI, and time spent reviewing its responses provided no detectable additional information about subsequent unaided performance beyond pre-test performance.

We then examined whether the within-session SRL and HS change scores provided information beyond these static characteristics. The temporal block comprised the two shift indices computed above to late change in SRL and HS activity. 

\begin{figure}[ht]
\centering
\includegraphics[width=1\textwidth, clip]{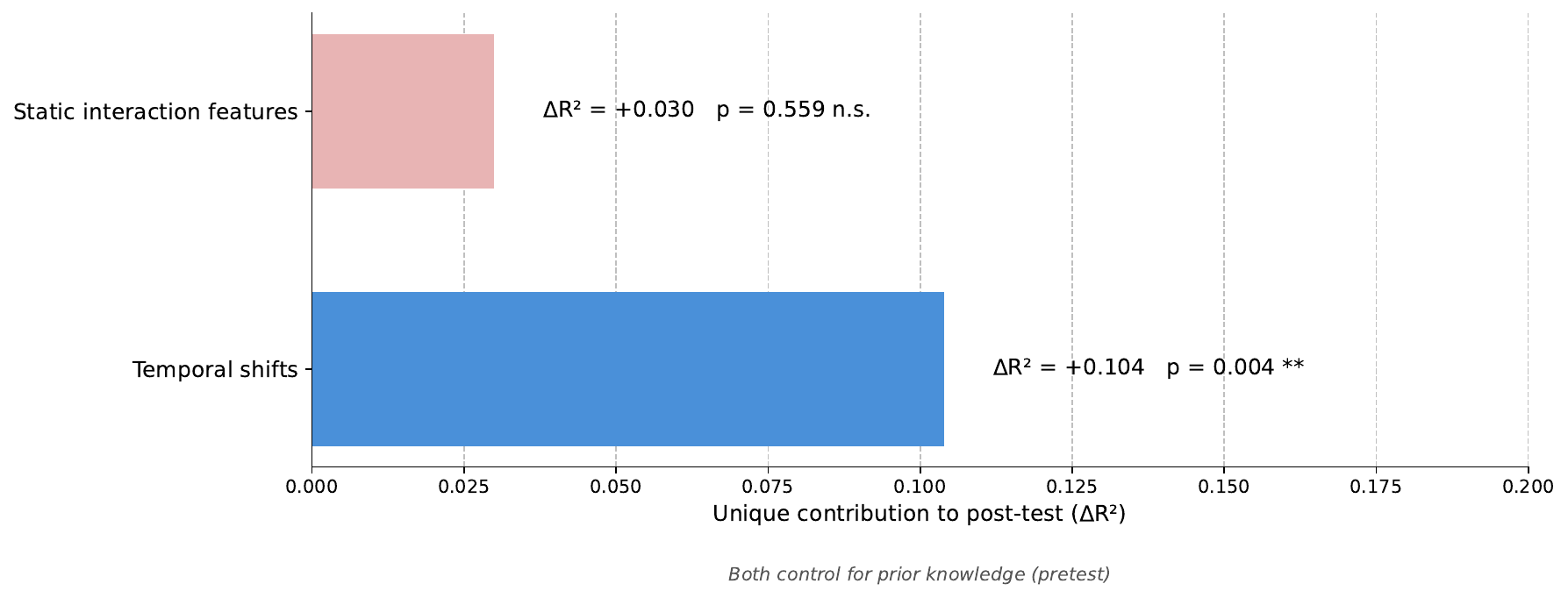}
\caption{Incremental and feature-level associations of static and dynamic interaction features with unaided post-test performance} 
\label{fig:post-test}
\end{figure}

Adding these features to pre-test performance alone increased explained variance from $R^2=.155$ to $R^2=.259$, $\Delta R^2=.104$, $F(2,85)=5.97$, $p=.004$. More importantly, after the static block had already been included, the temporal features increased explained variance from $R^2=.185$ to $R^2=.281$, $\Delta R^2=.096$, $F(2,81)=5.43$, $p=.006$. The reverse comparison did not show a corresponding contribution of the static features: after pre-test performance and the temporal block were entered, adding the static features increased explained variance only from $R^2=.259$ to $R^2=.281$, $\Delta R^2=.022$, $F(4,81)=0.62$, $p=.649$. Thus, the temporal features contained information about subsequent unaided performance that was not captured by students' aggregate interaction patterns.

This incremental association was primarily attributable to the development of students' HS. In the full model, the HS change score was positively associated with unaided post-test performance, $\beta=.29$, 95\% CI $[.10,.49]$, $p=.003$ (adjusted $p=.007$). Students whose requests became relatively more understanding-oriented---toward conceptual and procedural assistance rather than answer- and verification-seeking---subsequently performed better without AI assistance. Notably, this individual-level association occurred against the opposite cohort-level tendency: HS became more closure-oriented on average across the session ($M=-.19$, $t(88)=-2.69$, $p=.008$).

No association was detected for the SRL change score, $\beta=.14$, 95\% CI $[-.05,.33]$, $p=.155$. This null estimate should nevertheless be interpreted cautiously given the limited individual variation in the SRL change measure.

These results distinguish students' overall use of the AI from how their help-seeking developed--a relevant indicator for epistemic proactivity during learning--- during the interaction. The amount of understanding-oriented HS observed across the session was not by itself informative about subsequent unaided performance; rather, the relevant signal was whether students' requests became relatively more or less understanding-oriented as the interaction progressed.

\section{Discussion and Practical Implications}
This study examined how Grade-9 students regulated their use of a general-purpose LLM during open-ended mathematical-modelling practice, and whether their interaction behavior was associated with subsequent unaided performance. Three findings stand out. First, students' interactions were strongly request-dominated, with comparatively little explicit planning, monitoring, or evaluation, and their stated HS intentions were only weakly reflected in the assistance they subsequently sought. Second, students' behavior was not static: while explicit regulation increased modestly over the session, HS shifted on average toward more answer- and verification-seeking. Third, aggregate characteristics of AI use---including the overall prevalence of explicit regulation and understanding-oriented HS, as well as interaction duration and AI-response review time---did not account for detectable additional variance in unaided post-test performance beyond pre-test performance. In contrast, individual differences in HS development were informative: students whose requests became relatively more understanding-oriented (conceptual and procedural), rather than increasingly oriented toward obtaining or confirming outcomes, subsequently performed better without AI assistance.

Taken together, these patterns suggest that many students did not spontaneously organize their information-seeking with the LLM in a way that was consistently aligned with later independent performance. Although most entered the interaction with ostensibly learning-oriented intentions and frequently asked conceptual or procedural questions, their enacted strategies often diverged from those intentions and, on average, became more closure-oriented as the session progressed. This should not necessarily be interpreted as a stable limitation of the learners. The activity was deliberately open-ended: students were given a learning objective but little guidance about how to structure their interaction with the AI, and such self-directed use of a conversational system may itself be unfamiliar to secondary-school students. The observed general lack of SRL activities other than REQUEST could therefore partly reflect the regulatory demands created by the task format. The misalignment between students stated intentions are actual enacted behaviors could also reflect a change in their epistemic goals as their developed an understanding of the task, and had more experience with the system.

Our findings also suggest that the overall composition of a conversation can provide an incomplete picture of students' AI use. At the aggregate level, HS was relatively balanced, with slightly more conceptual and procedural than answer- and verification-seeking requests. Yet this summary obscured how assistance seeking changed as the interaction progressed. Indeed, the overall proportion of understanding-oriented HS was not detectably informative about subsequent performance, whereas its early-to-late change was. Analyses of student--AI interaction should therefore not treat prompts as independent indicators of productive AI use. The same request may have a different significance depending on the broader interaction in which it occurs: an occasional verification request may reflect appropriate checking, whereas an increasing reliance on the AI to provide or confirm outcomes may indicate greater delegation of epistemic work. Consequently, individual behaviors cannot be classified as inherently proactive or reactive outside their functional and temporal context. 

This distinction also clarifies how we interpret epistemic proactivity and responsibility during AI-supported learning. Our framework does not assume that learners should perform every cognitive or regulatory operation independently, nor that requesting answers or verification is intrinsically undesirable. Appropriate assistance can reduce unnecessary demands, particularly when learners are still developing knowledge of the task, and verification can constitute productive monitoring or calibration~\cite{Seufert2018TheLoad}. The principle underlying EP is instead that learners retain, or subsequently reclaim, responsibility for the learning-relevant epistemic work. In other words, delegation becomes potentially problematic when the learner increasingly relies on the model not only for support, but also for determining what to pursue, generating the relevant reasoning, and judging whether sufficient understanding has been reached.

This distinction is particularly important with general-purpose LLMs because the system can readily perform many of these operations and continuously propose how the interaction should proceed. Once learners experience such delegation as an efficient way to make progress, further offloading may become increasingly attractive~\cite{gerlich2025ai}. The average shift toward more closure-oriented HS observed in our sample is consistent with this possibility, although our data cannot establish why the shift occurred. Recent accounts of learning with AI similarly emphasize that productive use depends on how epistemic agency is distributed between learner and system, rather than on maximizing or minimizing AI assistance per se~\cite{yan2026agentivism}.

For AI-supported learning environments, these findings suggest that support based only on isolated prompts or whole-session usage statistics may miss potentially important changes in how learners are managing their learning process. A tutor might instead consider recent interaction history when deciding whether to intervene. For example, repeated or increasingly frequent requests for the AI to provide or confirm outcomes could motivate a scaffold that returns some of the epistemic work to the learner---by asking them to articulate their current reasoning, identify what remains unclear, or attempt the next step before receiving further assistance. Crucially, the goal would be to distinguish occasional, purposeful offloading from a developing pattern in which the AI increasingly performs the reasoning or evaluation that the learner is expected to practice. Such trajectory-sensitive scaffolding remains a design hypothesis to be experimentally tested rather than an intervention demonstrated by the present study.

The same caution applies to classroom implications. Our findings indicate that students may enter an AI-supported activity with learning-oriented intentions yet enact substantially different patterns of assistance seeking. This divergence may partly arise from the interactional affordances of general-purpose LLMs: their procedural fluency in producing explanations and suggested next steps can readily shape how an interaction unfolds and reduce the learners' role to determine the next epistemic move. Teachers may therefore find it useful to make the distribution of EP an explicit object of instruction: which parts of a task students want AI to support, which processes they are expected to practice themselves, and how they can recognize when assistance has begun to substitute for those processes. One possibility is to introduce think-first'' routines in which students articulate an initial interpretation, strategy, or uncertainty before consulting the AI, and then use the model to extend, challenge, or critique their reasoning. Students could also compare different ways of sequencing AI assistance---for example, immediately asking the AI to solve or explain a problem versus first developing their own reasoning and subsequently asking the AI for feedback---and reflect on what they understood and could subsequently do independently. Such activities could help make the consequences of different distributions of epistemic responsibility visible to learners rather than treating productive AI use as a matter of prompt formulation alone.

\section{Limitations, Future Directions and Conclusion}
Several limitations qualify these findings. First, the study used brief pre- and post-tests containing different but curriculum-matched MM items. Because each assessment contained only a small number of items, measurement error and possible differences in difficulty between the two forms limit the precision with which change in competence can be inferred. We therefore interpret the outcome as subsequent unaided post-test performance adjusted for pre-test performance, rather than as a direct measure of learning gain.

Second, the conversational measures are necessarily imperfect approximations of students' underlying regulatory processes. Agreement with expert-resolved labels was substantial for the collapsed contrasts used in the temporal analyses, but classification remained imperfect. The initially developed MM dimension showed insufficient automated reliability ($\kappa=.41$) and was therefore excluded from subsequent analyses. This difficulty may partly reflect the fit between traditional modeling-cycle categories and the turn-level analysis of AI-mediated learning. Future work should therefore investigate representations of modeling activity specifically suited to conversational AI.

Third, individual conversations were relatively short, providing only a small number of observations from which to estimate within-session development. Relatedly, the present analyses focused primarily on student turns. Because students' behavior develops in response to the content and structure of AI replies, future work should jointly model learner and assistant behavior to examine how particular system responses invite, constrain, or redirect subsequent help-seeking and regulation.

Fourth, the study examined a single interaction with one general-purpose LLM in one instructional context. It therefore remains unclear whether the observed patterns generalize across models, repeated use, different domains, or different forms of classroom orchestration. Longitudinal studies are particularly important for determining whether students' patterns of epistemic responsibility stabilize, deteriorate, or improve with experience.

Finally, the sample was drawn from three German Gymnasium schools and therefore represents a relatively specific academic-track population. Generalization to students in other school forms, educational systems, age groups, or levels of prior attainment should be made cautiously. 

Despite these limitations, the study provides first evidence that students' use of an LLM cannot be characterized adequately from isolated prompts or aggregate usage patterns alone. Students' HS changed meaningfully within a single interaction, and the direction of this change provided information about subsequent unaided performance that was not captured by how they used the system on average. Future work should test whether these patterns replicate across learners and contexts and, critically, whether trajectory-sensitive scaffolds can help students retain or reclaim EP.

\begin{credits}
\subsubsection{\ackname} Authors used GenAI tools during the editing and reviewing of the paper. Authors assume full responsibility for the content of this paper.

Work presented in this study benefited from funding from the LEAD Graduate School and Research Network for empirical education science at the University of Tübingen. 

\subsubsection{\discintname} Authors declare that the research was conducted in the absence of any relationship that could be construed as a potential conflict of interest.

\end{credits}

\bibliographystyle{splncs04}
\bibliography{references_lncs_corrected}

@techreport{shaw2026thinking,
  author = {Shaw, Steven D and Nave, Gideon},
  title = {Thinking--Fast, Slow, and Artificial: How {AI} Is Reshaping Human Reasoning and the Rise of Cognitive Surrender},
  institution = {The Wharton School, University of Pennsylvania},
  year = {2026},
  doi = {10.31234/osf.io/yk25n_v1},
  url = {https://ssrn.com/abstract=6097646},
  note = {The Wharton School Research Paper; available at SSRN 6097646}
}

@article{bulut2026ai,
  author = {Bulut, Neslihan and Borromeo Ferri, Rita},
  title = {{AI}-powered learning: exploring secondary school students' use of {ChatGPT} for solving mathematical modelling problems},
  journal = {ZDM -- Mathematics Education},
  year = {2026},
  doi = {10.1007/s11858-026-01803-6}
}

@inproceedings{ferri2010mathematical,
  author = {Borromeo Ferri, Rita and Blum, Werner},
  title = {Mathematical Modelling in Teacher Education---Experiences from a Modelling Seminar},
  booktitle = {Proceedings of the sixth Congress of the European Society for Research in Mathematics Education},
  pages = {2046--2055},
  year = {2010}
}

@inproceedings{alamray2026exploring,
  author = {Alamray, Fawzia and Aljohani, Naif R and Alfakeeh, Ahmed S and Alhebaishi, Nawaf and Ga{\v{s}}evi{\'c}, Dragan},
  title = {Exploring Students’ Cognitive Engagement with Generative Artificial Intelligence},
  booktitle = {Proceedings of the LAK26: 16th International Learning Analytics and Knowledge Conference},
  pages = {709--720},
  publisher = {Association for Computing Machinery},
  year = {2026},
  doi = {10.1145/3785022.3785132}
}

@inproceedings{liu2026behavioral,
  author = {Liu, Chang and Zhou, Qinyi and Shen, Xinjie and Liu, Xingyu Bruce and Wu, Tongshuang and Chen, Xiang'Anthony},
  title = {Behavioral Indicators of Overreliance During Interaction with Conversational Language Models},
  booktitle = {Proceedings of the 2026 CHI Conference on Human Factors in Computing Systems},
  pages = {1--23},
  publisher = {Association for Computing Machinery},
  year = {2026},
  doi = {10.1145/3772318.3790332}
}

@misc{yan2026agentivism,
  author = {Yan, Lixiang and Ga{\v{s}}evi{\'c}, Dragan},
  title = {Agentivism: a learning theory for the age of artificial intelligence},
  year = {2026},
  eprint = {2604.07813},
  archiveprefix = {arXiv},
  url = {https://arxiv.org/abs/2604.07813},
  note = {arXiv:2604.07813}
}

@article{gerlich2025ai,
  author = {Gerlich, Michael},
  title = {{AI} tools in society: Impacts on cognitive offloading and the future of critical thinking},
  journal = {Societies},
  volume = {15},
  number = {1},
  pages = {6},
  year = {2025},
  doi = {10.3390/soc15010006}
}

@article{Seufert2018TheLoad,
  author = {Seufert, Tina},
  title = {{The interplay between self-regulation in learning and cognitive load}},
  journal = {Educational Research Review},
  volume = {24},
  pages = {116--129},
  year = {2018},
  month = {6},
  doi = {10.1016/j.edurev.2018.03.004}
}

@article{maheux2026generative,
  author = {Maheux, Anne J and Akre-Bhide, Samir and Boeldt, Debra and Flannery, Jessica E and Richardson, Zachary and Burnell, Kaitlyn and Telzer, Eva H and Kollins, Scott H},
  title = {Generative Artificial Intelligence Applications Use Among US Youth},
  journal = {JAMA Network Open},
  volume = {9},
  number = {2},
  pages = {e2556631},
  year = {2026},
  doi = {10.1001/jamanetworkopen.2025.56631}
}

@article{kasneci2023,
  author = {Kasneci, Enkelejda and Se{\ss}ler, Kathrin and K{\"u}chemann, Stefan and Bannert, Maria and Dementieva, Daryna and Fischer, Frank and Gasser, Urs and Groh, Georg and G{\"u}nnemann, Stephan and H{\"u}llermeier, Eyke and Krusche, Stephan and Kutyniok, Gitta and Michaeli, Tilman and Nerdel, Claudia and Pfeffer, J{\"u}rgen and Poquet, Oleksandra and Sailer, Michael and Schmidt, Albrecht and Seidel, Tina and Stadler, Matthias and Weller, Jochen and Kuhn, Jochen and Kasneci, Gjergji},
  title = {{ChatGPT} for good? On opportunities and challenges of large language models for education},
  journal = {Learning and Individual Differences},
  volume = {103},
  pages = {102274},
  year = {2023},
  doi = {10.1016/j.lindif.2023.102274}
}

@article{tawfik2020role,
  author = {Tawfik, Andrew A and Graesser, Arthur and Gatewood, Jessica and Gishbaugher, Jaclyn},
  title = {Role of questions in inquiry-based instruction: Towards a design taxonomy for question-asking and implications for design},
  journal = {Educational Technology Research and Development},
  volume = {68},
  number = {2},
  pages = {653--678},
  year = {2020},
  doi = {10.1007/s11423-020-09738-9}
}

@book{holmes2023guidance,
  author = {Miao, Fengchun and Holmes, Wayne},
  title = {Guidance for Generative {AI} in Education and Research},
  publisher = {UNESCO},
  address = {Paris},
  year = {2023},
  doi = {10.54675/EWZM9535},
  isbn = {978-92-3-100612-8}
}

@article{blum2009mathematical,
  author = {Blum, Werner and Borromeo Ferri, Rita},
  title = {Mathematical modelling: Can it be taught and learnt},
  journal = {Journal of mathematical modelling and application},
  volume = {1},
  number = {1},
  pages = {45--58},
  year = {2009}
}

@article{Zimmerman2008InvestigatingProspects,
  author = {Zimmerman, Barry J.},
  title = {{Investigating self-regulation and motivation: Historical background, methodological developments, and future prospects}},
  journal = {American Educational Research Journal},
  volume = {45},
  number = {1},
  pages = {166--183},
  year = {2008},
  doi = {10.3102/0002831207312909}
}

@techreport{nelson1986help,
  author = {Nelson-Le Gall, Sharon},
  title = {Help-Seeking Behavior in Learning},
  number = {LRDC-1986/3},
  institution = {Learning Research and Development Center, University of Pittsburgh},
  year = {1986},
  month = mar,
  url = {https://eric.ed.gov/?id=ED275741},
  note = {ERIC ED275741}
}

@article{graesser1994question,
  author = {Graesser, Arthur C and Person, Natalie K},
  title = {Question asking during tutoring},
  journal = {American Educational Research Journal},
  volume = {31},
  number = {1},
  pages = {104--137},
  year = {1994},
  doi = {10.3102/00028312031001104}
}

@misc{abdelghani2025investigating,
  author = {Abdelghani, Rania and Murayama, Kou and Kidd, Celeste and Sauz{\'e}on, H{\'e}l{\`e}ne and Oudeyer, Pierre-Yves},
  title = {Investigating middle school students question-asking and answer-evaluation skills when using {ChatGPT} for science investigation},
  year = {2025},
  eprint = {2505.01106},
  archiveprefix = {arXiv},
  url = {https://arxiv.org/abs/2505.01106},
  note = {arXiv:2505.01106}
}

@techreport{hashem2025understanding,
  author = {Hashem, Youmna and Esnaashari, Saba and Onslow, Kate and Chakraborty, Sukankana and Francis, John and Poletaev, Anton and Bright, Jonathan},
  title = {Understanding the Impacts of Generative {AI} Use on Children: Work Package 1: Surveys},
  institution = {The Alan Turing Institute},
  year = {2025},
  url = {https://www.turing.ac.uk/news/publications/understanding-impacts-generative-ai-use-children-work-package-1-surveys}
}

@article{ruggeri2016sources,
  author = {Ruggeri, Azzurra and Lombrozo, Tania and Griffiths, Thomas L and Xu, Fei},
  title = {Sources of developmental change in the efficiency of information search.},
  journal = {Developmental Psychology},
  volume = {52},
  number = {12},
  pages = {2159--2173},
  year = {2016},
  doi = {10.1037/dev0000240}
}

@article{Yan2025DistinguishingAI,
  author = {Yan, Lixiang and Greiff, Samuel and Lodge, Jason M. and Ga{\v{s}}evi{\'{c}}, Dragan},
  title = {Distinguishing Performance Gains from Learning When Using Generative {AI}},
  journal = {Nature Reviews Psychology},
  volume = {4},
  pages = {435--436},
  year = {2025},
  doi = {10.1038/s44159-025-00467-5}
}

@article{stadler2024cognitive,
  author = {Stadler, Matthias and Bannert, Maria and Sailer, Michael},
  title = {Cognitive ease at a cost: {LLMs} reduce mental effort but compromise depth in student scientific inquiry},
  journal = {Computers in Human Behavior},
  volume = {160},
  pages = {108386},
  year = {2024},
  doi = {10.1016/j.chb.2024.108386}
}

\end{document}